\documentclass[%
 reprint,
showpacs,preprintnumbers,
 amsmath,amssymb,
 aps,
prc,
superscriptaddress]{revtex4-1}

\usepackage{graphicx}% Include figure files
\usepackage{dcolumn}% Align table columns on decimal point
\usepackage{bm}% bold math

\usepackage{hyperref}
\usepackage{booktabs}

\usepackage[T1]{fontenc}

\begin{document}

%\preprint{APS/123-QED}

\title{Hypermassive quark cores}% Force line breaks with \\
%\thanks{A footnote to the article title}%

\author{Luiz L. Lopes}
\email{luiz\_kiske@yahoo.com.br}
\affiliation{Centro Federal de Educa\c c\~ao  Tecnol\'ogica de
  Minas Gerais Campus VIII, CEP 37.022-560, Varginha, MG, Brasil}
\author{Carline Biesdorf}
\author{D\'ebora P. Menezes}
\affiliation{%
 Departamento de Fisica, CFM - Universidade Federal de Santa Catarina;  C.P. 476, CEP 88.040-900, Florian\'opolis, SC, Brasil 
}%

%\date{\today}% It is always \today, today,

\begin{abstract}

 Using a quantum hadrodynamics (QHD) and  MIT based models we construct hybrid stars within the Maxwell criteria of hadron-quark phase transition.
%With an extended version of the MIT bag model
We are able to produce a hybrid star with maximum mass of 2.15$M_\odot$. Furthermore, a 2.03$M_\odot$ star with a quark core corresponding to more than $80\%$ of both, its total mass and radius, is also possible.

\end{abstract}

%\pacs{21.65.Qr, 12.39.Ki}% PACS, the Physics and Astronomy
% Classification Scheme.

\maketitle

\section{Introduction}

Since the quantum chromodynamics (QCD) phase diagram became widely accepted, the possibility that neutron stars could, in fact, contain both a hadronic and a quark phase, started to be considered \cite{ivanenko1965hypothesis,universe7080267}.  In the same way that asymptotic freedom allows matter to become deconfined when the density increases at low temperatures, a similar behavior can take place in the interior of neutron stars. 

As the density increases towards the core of the star, quarks can become more energetically favorable than baryons, and, ultimately, the core of a neutron star may be composed of deconfined quarks. If the entire star does not convert itself into a quark star, as suggested by the Bodmer-Witten conjecture \cite{bodmer1971collapsed,witten1984cosmic}, the final composition is a quark core surrounded by a hadronic layer. This is what is generally called a hybrid star.

In the present work we construct hybrid stars using a quantum hadrodynamics (QHD) based model, the so called Walecka Model \cite{NLWM} with non-linear terms \cite{boguta1977relativistic}, to construct the hadronic equation of state (EoS) and MIT based models to construct the quark EoS. As for the MIT based models, we use the original MIT bag model, as introduced in ref.~\cite{chodos1974new,chodos1974baryon}, and explore two variations thereof, with a vector channel and with the vector channel altogether with the Dirac sea contribution, as introduced in ref.~\cite{lopes2021modified-partI,lopes2021modified-partII}.

For the hadronic model we use the parametrization presented in \cite{lopes2021hyperonic}, which fulfills most of the experimental constraints at the saturation density and also produces massive neutron stars, even when hyperons are present, fulfilling the constraints of the PSR J0740+6620~\cite{riley2021nicer}. For the quark model we use parametrizations that reproduce unstable strange matter, as we don't want the entire star to convert into a quark star, which could happen had we used stable strange matter.

We utilise the Maxwell construction to define the hadron-quark phase transition and to make the connection between the hadronic and the quark EoS. Finally, the TOV equations~\cite{oppenheimer1939massive} are solved to generate the macroscopic properties of the hybrid stars.

Even though various studies ~\cite{lopes2021broken,gerstung2020hyperon} show that hyperons seem inevitable, based on the fact that the hyperon onset in the stellar matter is still an open issue, we dedicate a section to study hybrid stars without hyperons.

We then use the Nambu Jona-Lasinio (NJL) model~\cite{nambu1961dynamical} to construct the quark EoS and make a comparison with the results obtained with the extended versions of the MIT bag model. We also investigate the differences between the hybrid stars with the parametrization of ref.~\cite{lopes2021hyperonic}, and with the traditional GM1~\cite{glendenning2000compact}, as presented in ref.~\cite{lopes2021broken}.

To finish, we analyze two important physical quantities, the dimensionless tidal parameter and the speed of sound. We compare the results for the tidal parameter to the recent constraints presented in the literature coming from LIGO/VIRGO gravitational wave observatories and the GW170817 event~\cite{abbott2018gw170817}. As for the speed of sound, we obtain its square value and study its relation with the size and mass of the quark core in hybrid stars as proposed in ref.~\cite{annala2020evidence}.

%%%%%%%%%%%%%%%%%%%%%%%%%%%%%%%%%%%%%%%%%%%%%%%%%%%%%%%%%%%%%%%%%%%%%%

\section{Hadronic EoS}\label{sec2}

The non-linear $\sigma\omega\rho$ QHD model has the following Lagrangian density \cite{glendenning2000compact}: 

\begin{eqnarray}
\mathcal{L}_{QHD} = \sum_B \bar{\psi}_B[\gamma^\mu(i\partial_\mu  - g_{B\omega}\omega_\mu   - g_{B\rho} \vec{\tau}_B \cdot \vec{\rho}_\mu)+ \nonumber \\
- (M_B - g_{B\sigma}\sigma)]\psi_B  -U(\sigma) +   \nonumber   \\
  + \frac{1}{2}(\partial_\mu \sigma \partial^\mu \sigma - m_s^2\sigma^2) - \frac{1}{4}\Omega^{\mu \nu}\Omega_{\mu \nu} + \frac{1}{2} m_v^2 \omega_\mu \omega^\mu+  \nonumber \\
 + \frac{1}{2} m_\rho^2 \vec{\rho}_\mu \cdot \vec{\rho}^{ \; \mu} - \frac{1}{4}\bf{P}^{\mu \nu} \cdot \bf{P}_{\mu \nu}  , \label{s1} 
\end{eqnarray}
in natural units. The sum on $B$ runs over the entire baryon octet: nucleons ($N$) and hyperons ($Y$). $\psi_B$  is the baryonic Dirac field. $\sigma$, $\omega_\mu$ and $\vec{\rho}_\mu$ are the mesonic fields. The $g's$ are the Yukawa coupling constants that simulate the strong interaction, $M_B$ is the baryon mass, $m_s$, $m_v$, and $m_\rho$ are
 the masses of the $\sigma$, $\omega$, and $\rho$ mesons, respectively. The antisymmetric mesonic field strength tensors are given by their usual expressions as presented in~\cite{glendenning2000compact}. $\vec{\tau}$ are the Pauli matrices. The $U(\sigma)$ is the self-interaction term introduced in ref.~\cite{boguta1977relativistic} to fix the incompressibility and is given by:
 
\begin{equation}
    U(\sigma) =  \frac{\kappa M_N(g_{\sigma} \sigma)^3}{3} + \frac{\lambda(g_{\sigma}\sigma)^4}{4}. \label{s2}
\end{equation}

Now, besides the traditional non-linear $\sigma\omega\rho$ QHD lagrangian density, we introduce two additional terms.  The first one is the strangeness hidden $\phi$ vector meson, which couples only with the hyperons ($Y$), not affecting the properties of symmetric matter:

\begin{equation}
\mathcal{L}_\phi = g_{Y \phi}\bar{\psi}_Y(\gamma^\mu\phi_\mu)\psi_Y + \frac{1}{2}m_\phi^2\phi_\mu\phi^\mu - \frac{1}{4}\Phi^{\mu\nu}\Phi_{\mu\nu}. \label{EL3} 
\end{equation}

\noindent as pointed out in ref.~\cite{lopes2021broken,cavagnoli2005importance,lopes2020role,weissenborn2012hyperons}, this vector channel is crucial to obtain massive hyperonic neutron stars. The second one is a non-linear  $\omega$-$\rho$ coupling term ~\cite{fattoyev2010relativistic}:

\begin{equation}
 \mathcal{L}_{\omega\rho} = \Lambda_{\omega\rho}(g_{N\rho}^2 \vec{\rho^\mu} \cdot \vec{\rho_\mu}) (g_{\omega}^2 \omega^\mu \omega_\mu) ,
\end{equation}

\noindent which is necessary to correct the slope of the symmetry energy ($L$) and has strongly influence on the radii and tidal deformation of the neutron stars~\cite{cavagnoli2011neutron,dexheimer2019we}.

In order to produce $\beta$ stable matter, with zero net charge, we also add leptons as a free Fermi gas.

The detailed calculations of the EoS for symmetric nuclear matter, as well as for $\beta$ stable matter in the QHD formalism are well documented and can be easily found in the literature~\cite{glendenning2000compact,serot1992quantum,universe7080267}.

%\begin{widetext}
\begin{center}
\begin{table}%[ht]
\begin{center}
\caption{Parameters of the model utilized in this work and their prediction for the symmetric nuclear matter properties at the saturation density; the phenomenological constraints are taken from ref.~\protect\cite{dutra2014relativistic,oertel2017equations}.}
\label{TL1}
\scalebox{0.90}{
\begin{tabular}{|c|c||c|c|c||c|}
\hline 
  & Parameters & &  Constraints  & This model  \\
 \hline
 $(g_{N\sigma}/m_s)^2$ & 12.108 $fm^2$ &$n_0$ ($fm^{-3}$) & 0.148 - 0.170 & 0.156 \\
 \hline
  $(g_{N\omega}/m_v)^2$ & 7.132  $fm^2$ & $M^{*}/M$ & 0.6 - 0.8 & 0.69  \\
  \hline
  $(g_{N\rho}/m_\rho)^2$ & 4.801  $fm^2$ & $K$ (MeV)& 220 - 260                                          &  256  \\
 \hline
$\kappa$ & 0.004138 & $S_0$ (MeV) & 28.6 - 34.4 &  31.2  \\
\hline
$\lambda$ &  -0.00390 & $L$ (MeV) & 36 - 86.8 & 74\\
\hline 
$\Lambda_{\omega\rho}$ &  0.0185 & $B/A$ (MeV) & 15.8 - 16.5  & 16.2  \\ 
\hline
\end{tabular}}
\end{center}
\end{table}
\end{center}
%\end{widetext}

The parametrization utilized in this work was retired from ref.~\cite{lopes2021hyperonic} and fulfills most of the experimental constraints at the saturation density and also produces massive neutron stars, even when hyperons are present, fulfilling the constraints of the PSR J0740+6620~\cite{riley2021nicer}. Its values, as well as the predictions of this model for the symmetric nuclear matter are presented in Table~\ref{TL1}. The nuclear constraints  at the saturation density are also in Table~\ref{TL1} and were taken from two extensive review articles, ref.~\cite{dutra2014relativistic,oertel2017equations}. Furthermore, the masses of the particles we use are the physical ones. The meson masses are $m_\omega$ = 783 MeV, $m_\rho$ = 770 MeV, $m_\phi$ = 1020 MeV, $m_\sigma$ = 512 MeV, the baryon octet masses are $M_N$ = 939 MeV, $M_\Lambda$ = 1116 MeV, $M_\Sigma$ = 1193 MeV, $M_\Xi$ = 1318 MeV, and the lepton masses are $m_e$ = 0.51 MeV, $m_\mu$ = 105.6 MeV.

When hyperons are present, it is crucial to define the strength of the hyperon-meson coupling constants. The only well known parameter is the $\Lambda^0$ hyperon potential depth, $U_\Lambda$ = -28 MeV. Here, we follow ref.~\cite{lopes2021hyperonic,lopes2014hypernuclear} and assume that both, vector and scalar mesons are constrained by SU(3) symmetry group~\cite{miyatsu2013equation,de1963octet}. In this case, given a value of $\alpha_v$, the value of $\alpha_s$ is automatically determined to reproduce $U_\Lambda$ = -28 MeV. In order to produce very stiff EoS we use here $\alpha_v$ = 0.50, which implies $\alpha_s$ = 0.911~\cite{lopes2021hyperonic}. Thereby, the ratios $g_{YB}/g_{NB}$ are~\cite{lopes2021broken,lopes2021hyperonic,lopes2014hypernuclear}:

\begin{align}
 g_{\Lambda\omega}/g_{N\omega} &= 0.714; \quad g_{\Lambda\phi}/g_{N\omega} = -0.808; \quad g_{\Lambda\rho}/g_{N\rho} = 0; \nonumber \\
 g_{\Sigma\omega}/g_{N\omega} &= 1.0; \quad g_{\Sigma\phi}/g_{N\omega} = -0.404 \quad g_{\Sigma\rho}/g_{N\rho} = 1.0; \nonumber \\
 g_{\Xi\omega}/g_{N\omega} &= 0.571; \quad g_{\Xi\phi}/g_{N\omega} = -1.01; \quad g_{\Xi\rho}/g_{N\rho} = 0; \nonumber \\
 g_{\Lambda\sigma}/g_{N\sigma} &= 0.646; \quad g_{\Sigma\sigma}/g_{N\sigma} = 0.690; \quad g_{\Xi\sigma}/g_{N\sigma} = 0.314; \nonumber \\
  \quad \quad
 \label{e5}
\end{align}

%%%%%%%%%%%%%%%%%%%%%%%%%%%%%%%%%%%%%%%%%%%%%%%%%%%%%%%%%%%%%%%%%%%%%%%%

\section{Quark EoS} \label{s3}

We compare here three variations of the MIT bag model. The original MIT bag model, as introduced in ref.~\cite{chodos1974new,chodos1974baryon}, and two variations, with a vector channel and with the vector channel altogether with the Dirac sea contribution, as introduced in ref.~\cite{lopes2021modified-partI,lopes2021modified-partII}.

\subsection{Original MIT bag model}

\begin{figure}%[ht] 
\begin{centering}
 \includegraphics[angle=270,
width=0.5\textwidth]{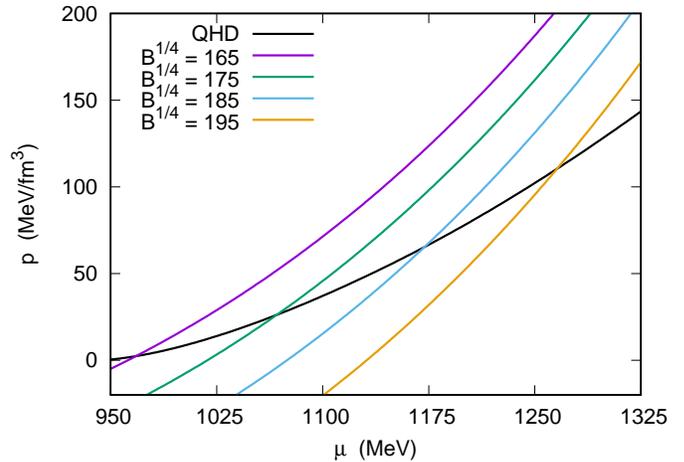}
\caption{(Color online) The pressure as a function of the chemical potential for different values of the Bag. The crossing points indicate the critical chemical potential, where the quark phase becomes energetically favorable. } \label{F1}
\end{centering}
\end{figure}

\begin{center}
\begin{table}%[ht]
\begin{center}
\caption{Critical chemical potential and pressure at the phase transition from hadron (H) to quark (Q) for different values of the Bag. We also show the energy density (in MeV/fm$^3$) at both phases at the critical neutron chemical potential.}
\label{T2}
\scalebox{0.99}{
\begin{tabular}{|c|c|c|c|c|c|}
\hline 
  B$^{1/4}$ (MeV)   &$\mu_n^H = \mu_n^Q$  & $p^H = p^Q$ & $\epsilon_H$ & $\epsilon_Q$ \\
 \hline
 165~ &  968 MeV & 2 (MeV$/fm^3)$ & ~121~ & ~407~ \\
 \hline
 175~ &  1067 MeV & 26 (MeV$/fm^3)$ & 319 & 596 \\
 \hline
 185~ &  1172 MeV  & 65 (MeV/$fm^3)$ & 440 & 832 \\
  \hline
 195~ & 1266 MeV  & 110 (MeV/$fm^3)$ & 558  & 1112 \\
  \hline
 \end{tabular}}
\end{center}
\end{table}
\end{center}

\begin{figure*}%[ht]
\begin{tabular}{cc}
\includegraphics[width=0.33\textwidth,angle=270]{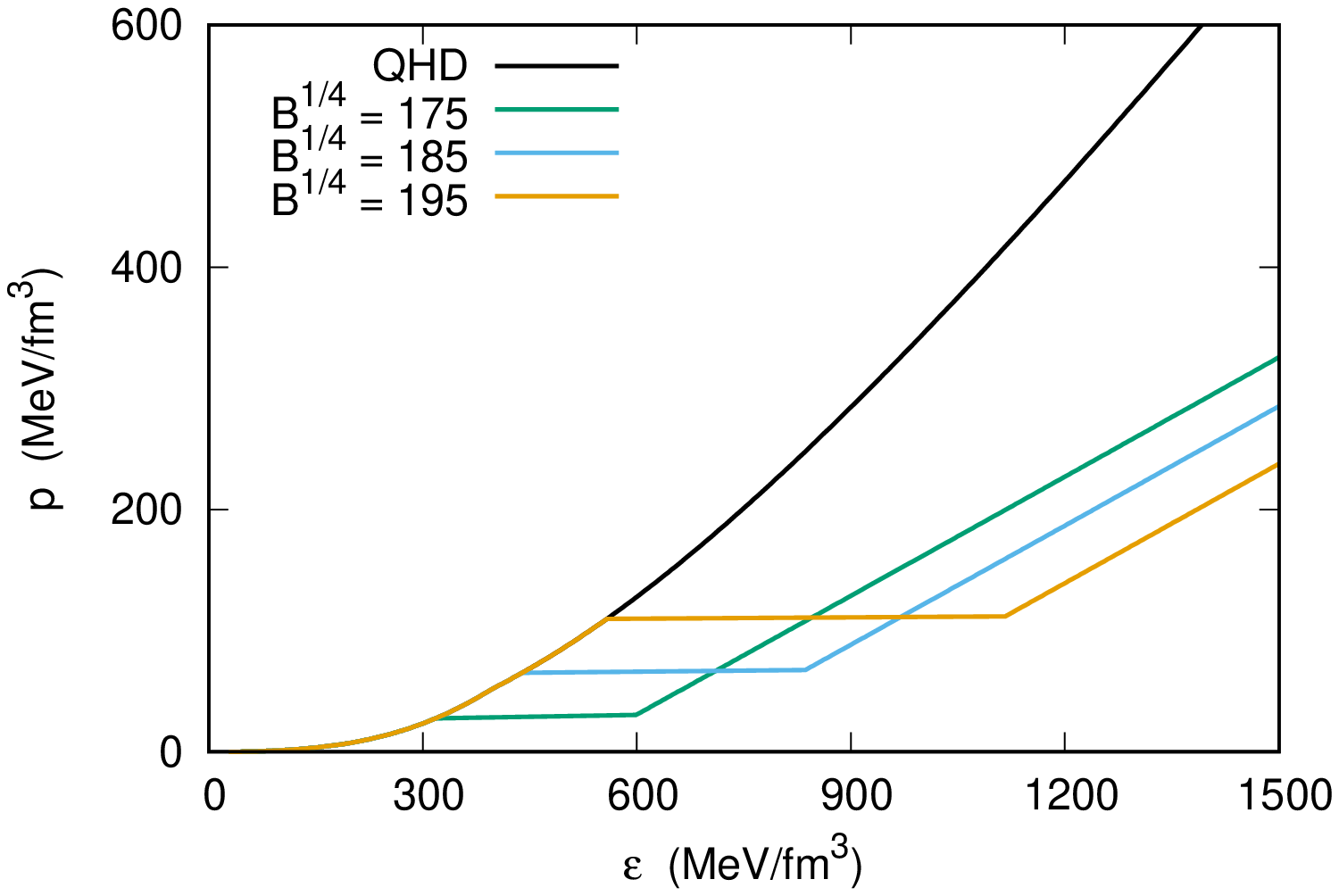} &
\includegraphics[width=0.33\textwidth,,angle=270]{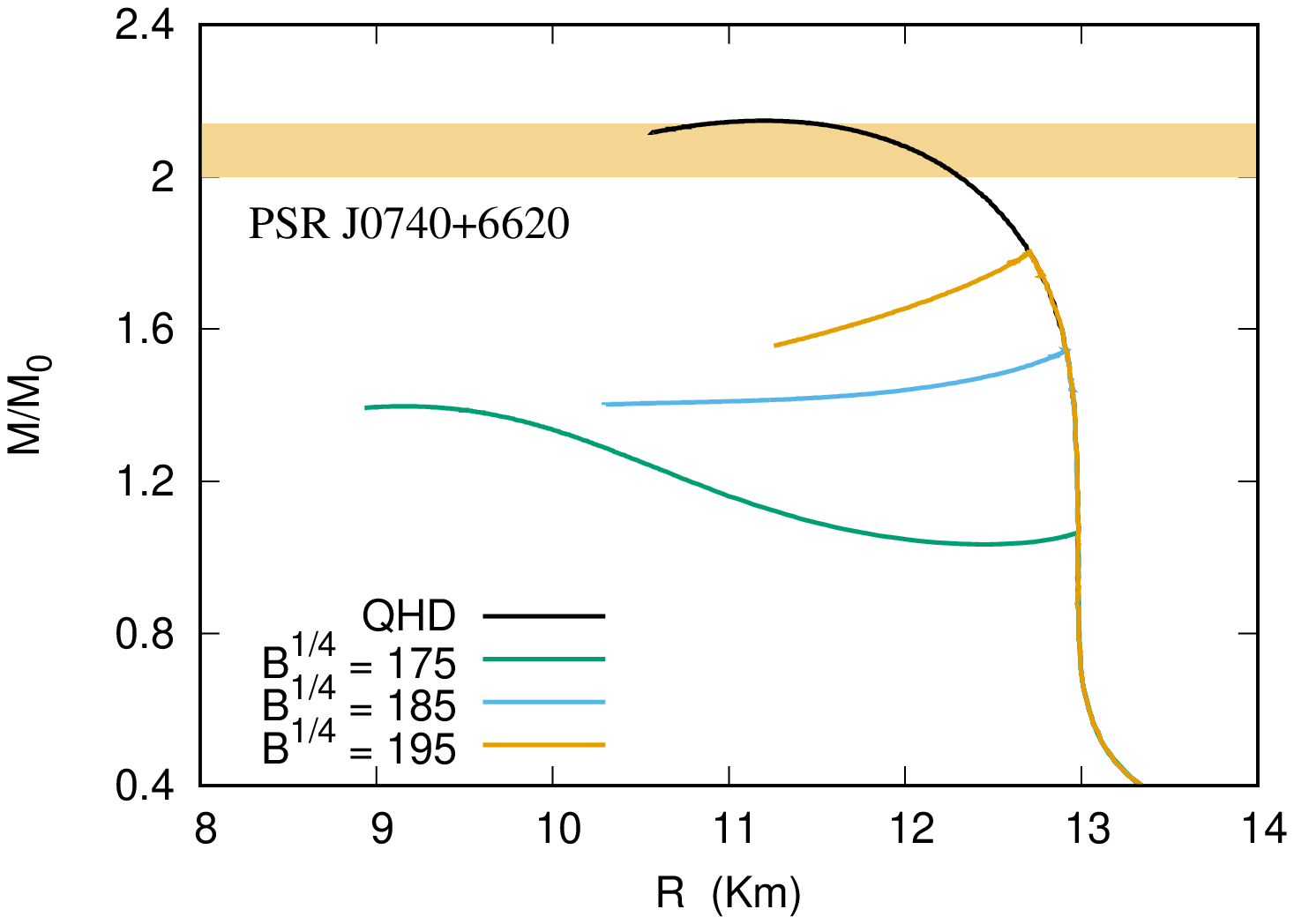} \\
\end{tabular}
\caption{(Color online) EoS (left) and the TOV solutions (right) for different values of the Bag. For the original Bag model no stable hybrid star reaches the mass of the PSR J070+6620.} \label{F2}
\end{figure*}

\begin{table}%[ht]
\begin{center}
\caption{Hybrid star properties for different values of the Bag. Only B$^{1/4}$ = 175 MeV produces stable hybrid stars. }\label{T3}
\scalebox{0.90}{
\begin{tabular}{|c|c|c|c|c|c|}
\hline
 Bag$^{1/4}$ (MeV) & $M/M_\odot$ & $ R~(km)$ & $\epsilon_c$  & $M_{min}/M_\odot$  & $ R_{1.4}~(km)$  \\
\hline
 175         & 1.40  & 9.15  &  2170   &  1.05  & 9.15\\
 \hline
185        & 1.54 & 12.90  & 444   & 1.54 & 12.96  \\
\hline
195       & 1.80 & 12.70  & 868   & 1.80 & 12.96  \\
\hline
QHD        & 2.15 & 11.20  & ~1279~   & - & 12.96  \\
\hline
\end{tabular}}
\end{center}
 \end{table}

The Lagrangian density of the original MIT bag model, as introduced in ref.~\cite{johnson1978field} reads:

\begin{equation}
 \mathcal{L_{MIT}} = \sum_{u,d,s} \{ \bar{\psi}_q[i\gamma^u\partial_u - m_q]\psi_q  - B \}\Theta(\bar{\psi}_q\psi_q), \label{e6}
\end{equation}
where $\psi_q$ is the Dirac field of the quark q, $m_q$ are the quark masses running over $u, d$ and $s$, whose values are 4 MeV, 4 MeV and 95 MeV, respectively~\cite{tanabashi2018review}. The Bag $B$ plays the role of the strong force, and it is the vacuum pressure constant. $\Theta(\bar{\psi}_q\psi_q)$ is the Heaviside step function to assure that the quarks exist only confined to the bag. The way to obtain the EoS in the original Bag model is well discussed in the ref.~\cite{lopes2021modified-partI,lopes2021modified-partII,johnson1978field}.

The free parameter $B$ is not completely arbitrary. If the $B$ value is too low, the $u-d$ quark matter is more stable than the hadronic matter, therefore, the universe as we know would not exist. If the $B$ is too high, both the $u-d$ matter and the $u-d-s$ quark matter, are unstable. However, for a small range of values, called stability window, the $u-d$ matter is unstable while the $u-d-s$ matter is stable. In this case the universe as we know is only meta-stable. This is the so called Bodmer-Witten conjecture. To more details see ref.~\cite{lopes2021modified-partI,lopes2021modified-partII} and the references therein. 

The values of the Bag pressure that fulfill the stability window are ultimately dependent on the $s$ quark mass. For the value used in this work, $m_s$ = 95 MeV, this corresponds to 148 MeV $<$ $B^{1/4}$ $<$ 159 MeV.

Now, as in this work we want to construct a hybrid star, we need to use an unstable $u-d-s$ matter. If the $u-d-s$ matter is stable, as soon as the core of the star converts to the quark phase, the entire star may convert into a quark star in a finite amount of time~\cite{olinto1987conversion}. Therefore, here we use $B^{1/4}~>~$ 159 MeV.

Another important quantity is the so called neutron critical chemical potential, the chemical potential value where the quark phase becomes energetically favorable, and a hadron-quark phase transition occurs. Its value is strongly model dependent, and it is defined at the point where the quark chemical potential equals the baryon chemical potential at the same pressure. Explicitly:

\begin{equation}
 \mu_n^H = \mu_n^Q \quad \mbox{and} \quad p_H = p_Q, \label{e7}
\end{equation}
where the neutron chemical potential can be written in terms of the quark ones as:

\begin{eqnarray}
 \mu_d = \mu_s = \frac{1}{3}(\mu_n + \mu_e) ,\nonumber \\
 \mu_u = \frac{1}{3} (\mu_n - 2\mu_e) . \label{e8}
\end{eqnarray}

The use of Eq.~\ref{e7} as the criteria to define hadron-quark phase transition is called Maxwell construction. There is no experimental evidence about the value of the neutron critical chemical potential at zero temperature. Therefore we follow ref.~\cite{lopes2021modified-partII} and only analyze models where the critical chemical potential lies between 1050 MeV $<~\mu_n~<$ 1400 MeV.

In Fig.~\ref{F1} we plot the pressure as a function of the chemical potential for the baryonic matter (QHD), as well as for different Bag pressure values in the original MIT bag model. When the QHD curve crosses the MIT one, we have the Maxwell criteria of  Eq.~\ref{e7}. The critical neutron chemical potential, as well as the pressure and the energy density (in both phases) are displayed in Table~\ref{T2}. As can be seen, for $B^{1/4}=165$~MeV, the critical chemical potential is below 1050 MeV, therefore, this result will not be analyzed any further. Notice that the critical chemical potential grows with the Bag pressure value. 

Now, we construct a hybrid EoS with different values of the Bag for the original MIT model. The EoS will be hadronic for values below the critical chemical potential, and a quark EoS is the preferential one for values above it. Then, we use this EoS as an input and solve the TOV equations~\cite{oppenheimer1939massive} to obtain the macroscopic properties of the hybrid stars. We also use the BPS EoS to simulate the neutron star crust~\cite{baym1971ground}. The results are presented in Fig.~\ref{F2}, and summarized in Table~\ref{T3}.

At first glance, it may seem that the maximum mass of a hybrid star grows with the Bag pressure value. But it is not what really happens. For values of $B^{1/4}~>$ 175 MeV, the central energy density lies in the window between the quark and hadron EoS. This region does not physically exist in the Maxwell construction, and the phases are spatially separated. Only with $B^{1/4}~$ = 175 MeV, we have a stable hybrid star with a maximum mass of 1.40 $M_\odot$, which is way below the observational limit of the PSR J070+6620~\cite{riley2021nicer}. Therefore, with the chosen QHD parametrization we cannot explain massive pulsars as a hybrid star within the original MIT bag model.

\subsection{Vector MIT bag model}

\begin{figure}%[ht] 
\begin{centering}
 \includegraphics[angle=270,
width=0.5\textwidth]{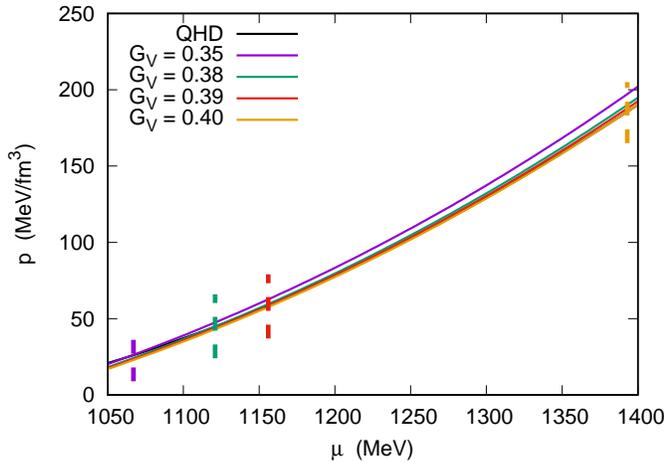}
\caption{(Color online) The pressure as a function of the chemical potential for different values of $G_V$. The dotted vertical lines indicate the position of the critical chemical potential.} \label{F3}
\end{centering}
\end{figure}

\begin{center}
\begin{table}%[ht]
\begin{center}
\caption{Chemical potential and pressure at phase transition for hadron (H) to quark (Q) for different values of $G_V$. We also show the energy density (in MeV/fm$^3$) at both phases at the critical neutron chemical potential.}
\label{T4}
\scalebox{0.99}{
\begin{tabular}{|c|c|c|c|c|c|}
\hline 
  $G_V$ (fm$^2$)   &$\mu_n^H = \mu_n^Q$  & $p^H = p^Q$ & $\epsilon_H$ & $\epsilon_Q$ \\
 \hline
 0.35~ &  1063 MeV & 25 (MeV$/fm^3)$ & ~303~ & ~364~\\
 \hline
 0.38~ &  1120 MeV & 45 (MeV$/fm^3)$ & 375 & 407 \\
 \hline
 0.39~ &  1156 MeV  & 58 (MeV/$fm^3)$ & 415 & 440 \\
  \hline
 0.40~ & 1394 MeV  & 186 (MeV/$fm^3)$ & 720  & 735 \\
  \hline
 \end{tabular}}
\end{center}
\end{table}
\end{center}

We can overcome the issue of not obtaining massive hybrid stars by introducing a vector channel to the original MIT bag model, similar to the vector channel in QHD models, as was done in ref.~\cite{lopes2021modified-partI,lopes2021modified-partII}. To this end, we add the following Lagrangian density to the original one presented in Eq.~\ref{e6}:

\begin{equation}
 \mathcal{L}_V = \{\sum_{u,d,s}\bar{\psi}_q g_{qV}\psi_q - \frac{1}{2}m_V^2 V^\mu V_\mu \} \Theta(\bar{\psi}_q\psi_q) ,\label{e9}
\end{equation}
where the quark interaction is mediated by the vector channel $V_\mu$, analogous to the $\omega$ meson in QHD. Indeed, in this work we consider that the vector channel is the $\omega$ meson itself. The inclusion of vector channels in the MIT bag model is not new, as can be seen in~\cite{gomes2019can,gomes2019constraining,franzon2016effects,klahn2015vector,wei2019camouflage}.

Unfortunately, in the previous papers, the authors missed the mass term of the vector channel. As discussed in ref.~\cite{lopes2021modified-partI}, in mean field approximation, the vector channel becomes zero if the mass is zero. The construction of the EoS within the vector MIT bag model in mean field approximation is analogous to the QHD one, and it is well discussed in ref.~\cite{lopes2021modified-partI,lopes2021modified-partII}.

It is interesting to notice that both the QHD and the vector MIT models use the exchange of massive mesons to simulate the strong force, indicating an unified interaction scheme. Moreover, as we know the quantum numbers related to the isospin, hypercharge for the baryons, quarks and mesons, we can also use the same criteria here as we used in the QHD to fix the relative strength of the quarks with the $\omega$ meson. In other words, we also use here the SU(3) symmetry group. The detailed calculation is presented in the appendix of ref.~\cite{lopes2021modified-partI}. We have:

\begin{equation}
g_{sV}/g_{uV} = g_{sV}/g_{dV} =  0.40. \label{e10}
\end{equation}

Another important quantity is the $G_V$ defined as:

\begin{equation}
 G_V~\dot{=}~(g_{uV}/m_V)^2 , \label{e11}
\end{equation}
 and is related to the strength of the vector field. Higher the value of $G_V$, higher the vector field interaction and stiffer is the EoS. It is clear from Eq.~\ref{e11} that, to keep thermodynamic consistence, the mass term cannot be zero, at least at mean field approximation. Also, in the vector MIT bag model, the stability window now depends on the value of the $G_V$. The higher the value of $G_V$, the lower the maximum value of $B$ that satisfies the Bodmer-Witten conjecture. Indeed, as shown in ref.~\cite{lopes2021modified-partI}, for $G_V$ = 0.3 fm$^2$, the stability window lies between 139 MeV $<~B^{1/4}~<$ 150 MeV. Here, we use $~B^{1/4}~$ = 158 MeV, ensuring an unstable $u-d-s$ matter.
 
 We plot the pressure as function of the chemical potential for the vector MIT bag model within different values of $G_V$ in Fig.~\ref{F3}. Unlike Fig.~\ref{F2} for the original MIT, here the crossing of the hadron-quark curves are not so easily identifiable. This is because the slope of the curves in both phases are very similar. To help identify the critical chemical potential, we plot dotted vertical lines in Fig.~\ref{F3}. The neutron critical chemical potential, the pressure and the energy density for each value of $G_V$ are displayed in Table~\ref{T4}. As can be seen, the higher the value of $G_V$, the higher the critical chemical potential. We also see that $G_V$ must lie between 0.35 fm$^2$ $<~G_V~<$ 0.40  fm$^2$ so that  the critical chemical potential falls between 1050 MeV $<~\mu~<$ 1400 MeV.
 
 \begin{figure*}%[ht]
\begin{tabular}{cc}
\includegraphics[width=0.33\textwidth,angle=270]{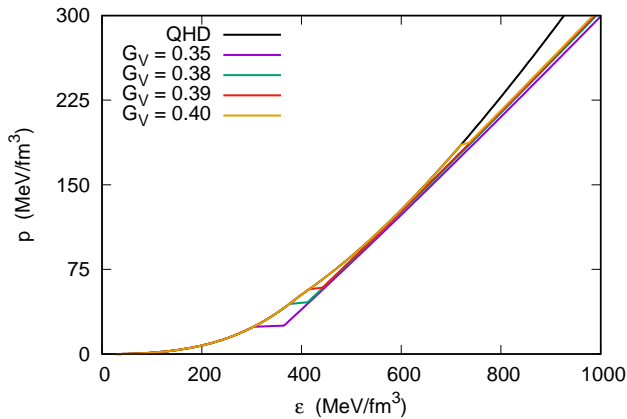} &
\includegraphics[width=0.33\textwidth,,angle=270]{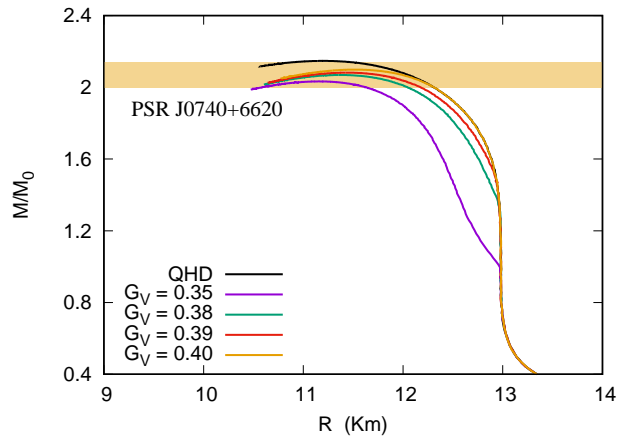} \\
\end{tabular}
\caption{(Color online) EoS (left) and TOV solutions (right) for different values of $G_V$.} \label{F4}
\end{figure*}

\begin{table}%[ht]
\begin{center}
\caption{Hybrid star properties for different values of the $G_V$.}
\label{T5}
\begin{tabular}{|c|c|c|c|c|c|}
\hline
$G_V$ (fm$^2$) & $M/M_\odot$ & $ R~(km)$ & $\epsilon_c$  & $M_{min}/M_\odot$  & $ R_{1.4}~(km)$  \\
\hline
  0.35    & 2.03  & 11.15  &  1282   &  0.98  & 12.54\\
 \hline
 0.38         & 2.07  & 11.36  &  1235   &  1.31  & 12.91\\
 \hline
0.39        & 2.08 & 11.44  & 1217   & 1.47 & 12.96  \\
\hline
0.40       & 2.10 & 11.51  & 1195   & 1.99 & 12.96  \\
\hline
QHD        & 2.15 & 11.20  & ~1279~   & - & 12.96  \\
\hline
\end{tabular}
\end{center}
 \end{table}

We plot in Fig.~\ref{F4} the EoS and the corresponding TOV solutions for the hybrid stars within the vector MIT bag model. Compared with the original MIT bag model, we see that the EoS are much stiffer within the vector channel. We also see that the gap in the energy density is significantly smaller here than in the original one. Another point is the fact that varying $G_V$ from 0.35 to 0.40 fm$^2$, although it causes a large effect on the value of the critical chemical potential, it does not affect too much the stiffness of the EoS. All these facts are reflected in the mass-radius relation. For instance, the maximum mass of a hybrid star varies from 2.03 $M_\odot$ to 2.10 $M_\odot$, a difference around $3\%$, but the $M_{min}$, the minimum mass value for which the hybrid star is formed, varies from 0.98 $M_\odot$ to 1.99 $M_\odot$, a difference greater than 100$\%$. Indeed, for $G_V$ up to values of 0.38 fm$^2$, the hadron-quark phase transition takes place below the hyperon threshold.
The first hyperon onset happens around $n = 0.42$ fm$^{-3}$, and at a neutron chemical potential of $\mu_n$ = 1131 MeV.  A critical chemical potential below 1131 MeV  can, therefore, be faced as a possible solution to the so-called hyperon puzzle. As displayed in Fig.~\ref{F4}, if the hadron-quark phase transition happens at low density, the radius of the canonical 1.4 $M_\odot$ can be significantly lower than its pure hadronic counter-part. All the main results are summarized in Table~\ref{T5}.

\begin{center}
\begin{table}%[ht]
\begin{center}
\caption{Masses and radii of the quark core and their proportional   contribution for the maximally massive star within different hybrid stars models.}
\label{T6}
\scalebox{0.99}{
\begin{tabular}{|c|c|c|c|c|c|c|c|}
\hline 
 $G_V$  (fm$^2)$  &  ~0.35~ &  ~0.38~  &  ~0.39~  &  ~0.40~  \\
 \hline
 $M_Q/M_\odot$  & 1.65  & 1.42 & 1.30 & 0.50   \\
 \hline
 R$_Q$  (km) & 9.06  & 8.44 & 8.08 & 5.30  \\
 \hline
 $\%$ $M_Q/M_{max}$  & 81$\%$  & 69$\%$ & 63$\%$  & 23$\%$  \\
 \hline
 $\%$  R$_Q$/R$_{total}$ & 81$\%$  & 74$\%$ & 70$\%$  & 46$\%$      \\
 \hline
\end{tabular}}
\end{center}
\end{table}
\end{center}

Now we confront our results in the light of a very recent study about the nature of massive stars. In ref.~\cite{annala2020evidence}, the authors suggest that massive neutron stars have sizable quark core. More than that, they argue that hybrid stars are not only possible, but also probable. They found that for a two solar masses star with a radius of 12 km, the quark core can reach 0.8 $M_\odot$ (40$\%$) and 7 km (58$\%$) of the total composition of the star.

We estimate the mass and size of the quark core in the most massive hybrid star of each model presented in Table~\ref{T5}. To accomplish that we solve the TOV equations for the quark EoS from the energy density corresponding to the critical pressure displayed in Table~\ref{T4} up to the energy density at the maximum mass shown in Table~\ref{T5}. The results are presented in absolute and relative values in Table~\ref{T6}. As can be seen, within the vector MIT bag model, the quark core can be even more significant than within the pQCD used in ref.~\cite{annala2020evidence}. Indeed, for $G_V$ = 0.35 fm$^2$, we are able to produce hybrid stars with a hypermassive quark core of 1.65 $M_\odot$ and a radius larger than 9 km. In relative values, this corresponds to more than 80$\%$ of both the total mass and radius of the hybrid star. It is also worthy to point out that such hybrid star with the hypermassive quark core still fulfills the constraints of the PSR J0740+6620~\cite{riley2021nicer}.

We can also compare our results with other hybrid stars studies found in the literature. For instance, our results are close to those presented in ref.~\cite{gomes2019constraining}, although both the hadronic and the quark models are different. Here, we use an extended version of the QHD, with the parametrization taken from ref.~\cite{lopes2021hyperonic}, while ref.~\cite{gomes2019constraining} uses a many-body model with a derivative coupling. Both models predict the same maximum mass of 2.15 $M_\odot$. The main macroscopic difference is the radius of the canonical star: 12.96 km here vs 14.44 km in ref.~\cite{gomes2019constraining}. Both studies use the vector MIT bag model, although in ref.~\cite{gomes2019constraining} the mass term, important to keep the thermodynamic consistency, is missing. Within the set with the larger quark core, in Maxwell construction, we found a maximum mass of 2.03 $M_\odot$ with a $M_{min}$ = 0.98 $M_\odot$, while ref.~\cite{gomes2019constraining} found a maximum mass of 1.99 $M_\odot$ with a $M_{min}$ = 1.03 $M_\odot$.

On ref.~\cite{klahn2015vector}, the authors use the so-called tdBAG, which is, ultimately, based on the NJL models for the quark phase, and the TM1 parametrization of the QHD for the hadron phase. They found that only with two flavor $u-d$ quark matter the phase transition is possible. And, unlike in the present work, as well as in ref.~\cite{gomes2019constraining}, the hybrid star has a very similar radius with respect to the pure hadronic one.

\subsection{Dirac sea contribution}

The vector channel in the mean field approximation takes into account only the valence quarks. This scenario is called `no sea approximation', once the Dirac sea of quarks is completely ignored~\cite{furnstahl1997vacuum}. As the vector field is borrowed directly from quantum hadrodynamics (QHD), the vector MIT bag model also becomes renormalizable~\cite{serot1992quantum}. However, instead of transforming the mean field approximation (MFA) into a more complex relativistic Hartree or Hartree–Fock approximation, we can take the Dirac sea into account throughout modifications on the effective Lagrangian density as done in~\cite{furnstahl1997vacuum}. Here, we follow ref.~\cite{lopes2021modified-partI}, and  introduce a quartic contribution for the vector field as a correction for the EoS at high density which will mimic the Dirac sea contribution. Therefore, we add the Lagrangian of Eq.~\ref{e12} to the Lagrangians of Eq.~\ref{e6} and Eq.~\ref{e5}.

\begin{equation}
 \mathcal{L}_{Dirac} = b_4\frac{(g^2V_\mu V^\mu)^2}{4} .\label{e12}
\end{equation}
where we define $g=g_{uV}$.

\begin{figure}%[ht] 
\begin{centering}
 \includegraphics[angle=270,
width=0.5\textwidth]{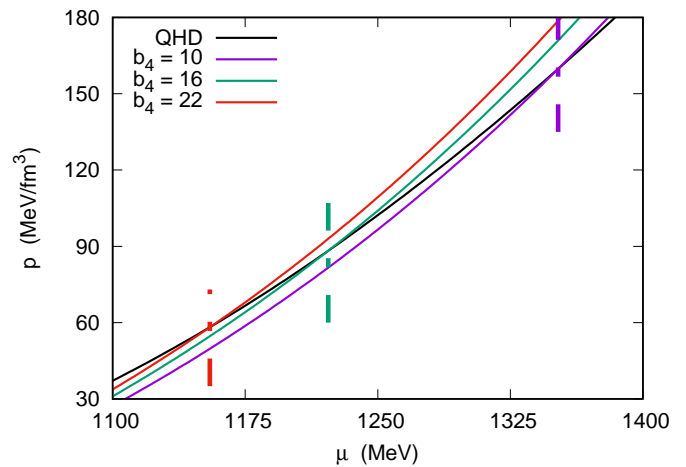}
\caption{(Color online) The pressure as a function of the chemical potential for different values of $b_4$ with $G_V$=0.8 fm$^2$. The dotted vertical lines indicate the position of the critical chemical potential.} \label{F5}
\end{centering}
\end{figure}

\begin{center}
\begin{table}%[ht]
\begin{center}
\caption{Chemical potential and pressure at phase transition for hadron (H) to quark (Q) for different values of $b_4$. We also show the energy density (in MeV/fm$^3$) at both phases at the critical chemical potential.}
\label{T7}
\scalebox{0.99}{
\begin{tabular}{|c|c|c|c|c|c|}
\hline 
  ~$b_4$~  &$\mu_n^H = \mu_n^Q$  & $p^H = p^Q$ & $\epsilon_H$ & $\epsilon_Q$ \\
 \hline
 10~ &  1352 MeV & 159 (MeV$/fm^3)$ & ~667~ & ~780~\\
 \hline
 16~ &  1222 MeV & 88 (MeV$/fm^3)$ & 502 & 579 \\
 \hline
 22~ &  1154 MeV  & 58 (MeV/$fm^3)$ & 415 & 493 \\
 \hline
 \end{tabular}}
\end{center}
\end{table}
\end{center}

As pointed out in ref.~\cite{lopes2021modified-partI,lopes2021modified-partII,furnstahl1997vacuum}, the quartic term introduced in Eq.~(\ref{e12}) softens the EoS at high density, which allows us to use higher values of $G_V$ without crossing the limit $\mu~=$ 1400 MeV. Indeed, in this section we use $G_V$ = 0.8 fm$^2$ and keep $B^{1/4}$ = 158 MeV, as in the previous one. The construction of an EoS with the term which mimics the Dirac sea contribution is discussed in ref.~\cite{lopes2021modified-partI,lopes2021modified-partII,furnstahl1997vacuum}. In Fig.~\ref{F5} we plot the pressure as a function of the chemical potential for three different values of $b_4$. The vertical dotted lines are also plotted to help identify the neutron critical chemical potential. The results are also shown in Table~\ref{T7}. As can be seen, there is an already expected relation between the critical chemical potential and the value of $b_4$. The higher the $b_4$ value, the softer is the quark EoS and therefore, the lower is the critical chemical potential. Even with the high value $G_V$ = 0.8 fm$^2$, we can choose values of $b_4$ able to produce the critical chemical potential in the range between 1050 MeV $<~\mu~<$ 1400 MeV.

\begin{figure*}%[ht]
\begin{tabular}{cc}
\includegraphics[width=0.33\textwidth,angle=270]{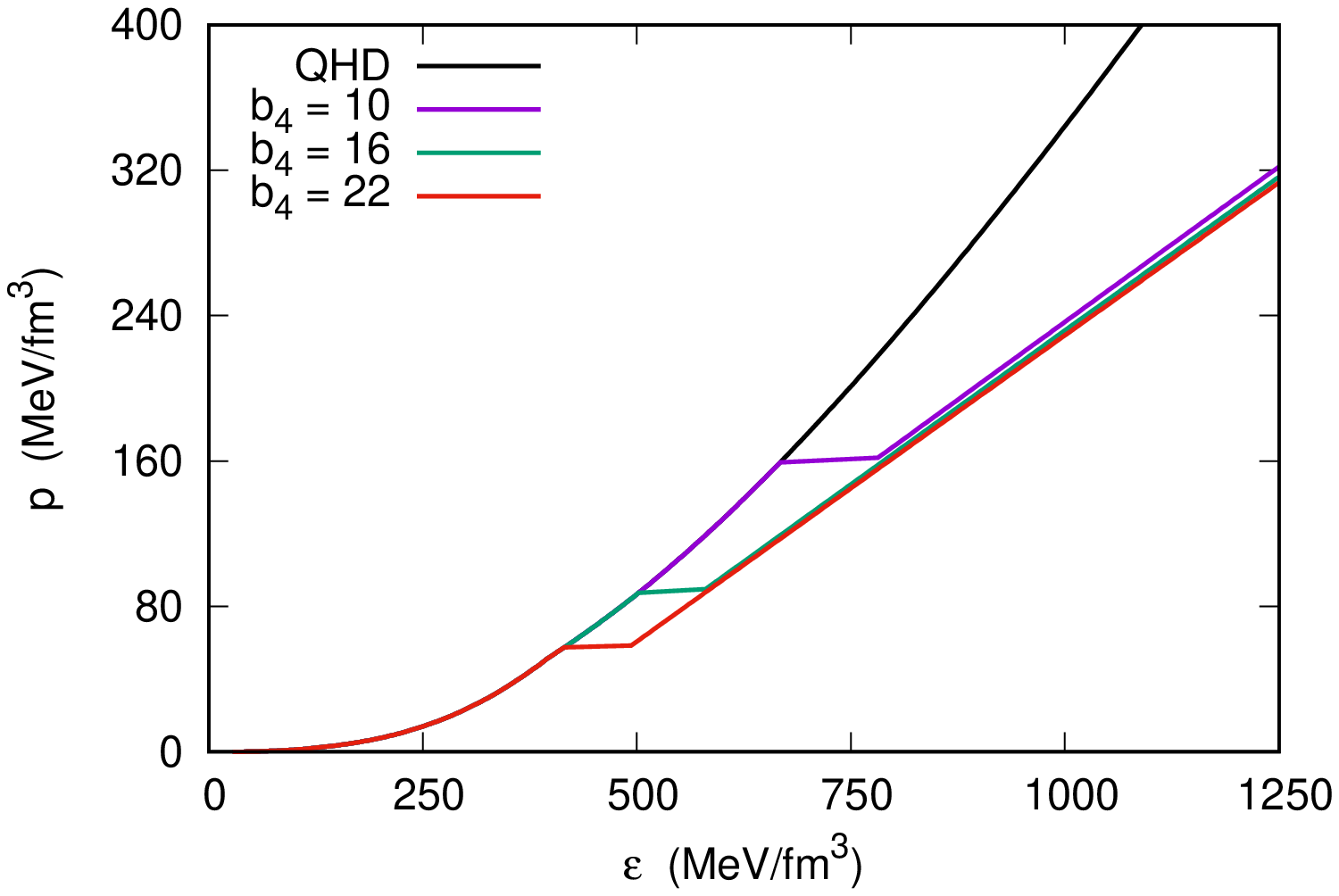} &
\includegraphics[width=0.33\textwidth,,angle=270]{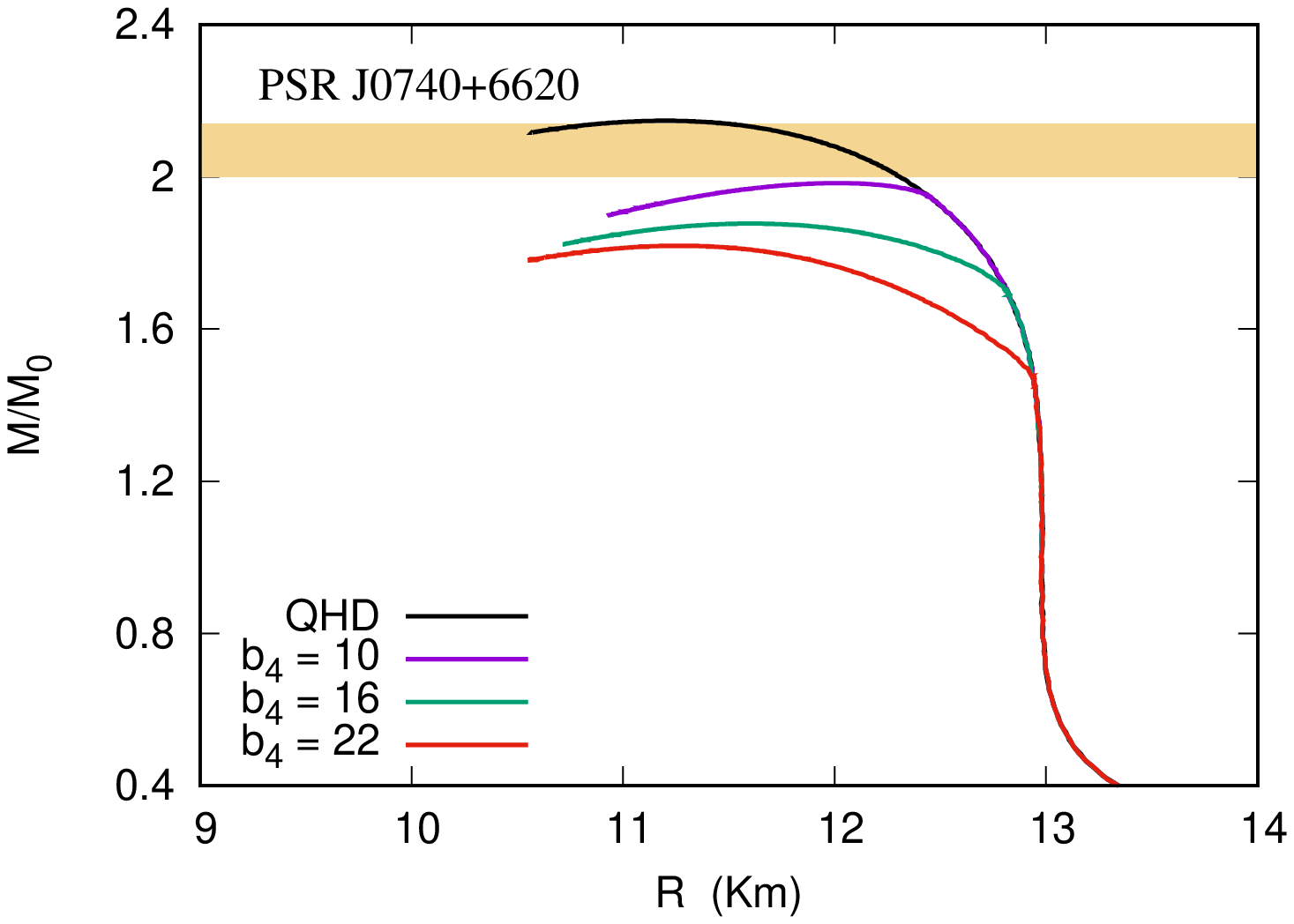} \\
\end{tabular}
\caption{(Color online) EoS (left) and TOV solutions (right) for different values of $b_4$. No hybrid star reaches the mass of the PSR J0740+6620, not even with very a high value of $G_V$ = 0.8 fm$^2$.} \label{F6}
\end{figure*}

\begin{table}%[ht]
\begin{center}
 \caption{Hybrid star properties for different values of the $b_4$ with $G_V$ = 0.8 fm$^2$.}\label{T8}
\begin{tabular}{|c|c|c|c|c|c|}
\hline
~$b_4$~& $M/M_\odot$ & $ R~(km)$ & $\epsilon_c$  & $M_{min}/M_\odot$  & $ R_{1.4}~(km)$  \\
\hline
  10   & 1.98  & 12.01  &  1035   &  1.95  & 12.96\\
 \hline
 16         & 1.88  & 11.60 &  1170   &  1.69  & 12.96\\
 \hline
22        & 1.82 & 11.26  & 1278   & 1.46 & 12.96  \\
\hline
QHD        & 2.15 & 11.20  & ~1279~   & - & 12.96  \\
\hline
\end{tabular}
\end{center}
 \end{table}

Now, in Fig.~\ref{F6} we plot the EoS and the TOV solutions for different values of $b_4$. The main results are also displayed in Table~\ref{T7}. As we can see, the Dirac sea contribution makes the gap in the energy density larger than in the previous section, without the quartic term. Also, we can see that the gap depends on $b_4$. The higher the value of $b_4$, the smaller the gap. With respect to the TOV solutions, we see that, despite the high value of $G_V$, with the quartic term in the quark phase, no hybrid star reaches the observed limit of the PSR J0740+6620~\cite{riley2021nicer}. Due to this drawback, we do not calculate the relative values of the quark core. It is, nevertheless, worth pointing out that, for all utilized values of $b_4$, we are able to produce a stable hybrid branch. Also, unlike the vector case, here, the maximum mass strongly depends on the value of $b_4$. The maximum mass can vary from 1.82 M$_\odot$ to 1.98 $M_\odot$, a difference of about $9\%$. Furthermore, within the Dirac sea approximation, no canonical star has a quark phase.

%%%%%%%%%%%%%%%%%%%%%%%%%%%%%%%%%%%%%%%%%%%%%%%%%%%%%%%%%%%%%%%%%%%%%%%%%%%%

\section{No hyperons}

One of the open issues in nuclear astrophysics is the content of the inner core of massive neutron stars. Because of the Pauli principle, as the number density increases, the  Fermi energy of the nucleons exceeds the mass of the heaviest baryons, and the conversion of some nucleons into hyperons becomes energetically favorable. It is also well known that the hyperon onset softens the EoS. In some cases, the softening of the EoS pushes the maximum mass below the observational limit of massive neutron stars. This possible conflict between theory and observation is called hyperon puzzle. 
Among others, two extensive studies about the hyperon threshold~\cite{lopes2021broken, dhapo2010appearance} show that hyperons are, ultimately, inevitable. This is the reason why, up to now, we have always considered the EoS with hyperons.

Nevertheless, as pointed out, the hyperon onset in stellar matter is still an open issue. Moreover, a recent study on non-relativistic models shows that the use of strongly repulsive three-body forces can suppress the hyperon threshold~\cite{gerstung2020hyperon}. Therefore we dedicate this section to study hybrid stars without hyperons. But we have to keep in mind that, for the QHD model we use in the present paper, the hyperon suppression 
is obtained simply by not taking them into account.

%can only be obtained artificially, i.e, we do not take them into account in the EoS, otherwise, they will always be present.

\begin{figure}%[ht] 
\begin{centering}
 \includegraphics[angle=270,
width=0.5\textwidth]{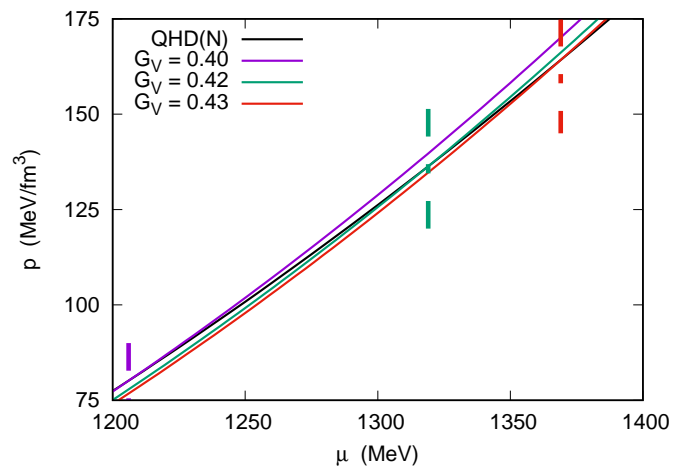}
\caption{(Color online) The pressure as a function of the chemical potential for different values of $G_V$ without hyperons. The dotted vertical lines indicate the position of the critical chemical potential.} \label{F7}
\end{centering}
\end{figure}

As the vector MIT bag model without the Dirac sea contribution predicts the hybrid stars with the highest maximum mass, as well as the biggest quark core, we only use this set in this section. Also, as discussed in previous sections, with $G_V~<$ 0.39 fm$^2$, the hadron-quark phase transition happens before the hyperon onset. Therefore, it would not make any difference to use these values. So, we restrict our analyzes next to values of $G_V~>$ 0.39. We plot in Fig.~\ref{F7} the pressure as a function of the chemical potential. We also plot the dotted lines to help the identification of the critical chemical potential. The results are also presented in Table~\ref{T9}. We can see that, without hyperons, we can use higher values of $G_V$ while keeping the critical chemical potential in between 1050 MeV $<~\mu~<$ 1400 MeV.

\begin{figure*}%[ht]
\begin{tabular}{cc}
\includegraphics[width=0.33\textwidth,angle=270]{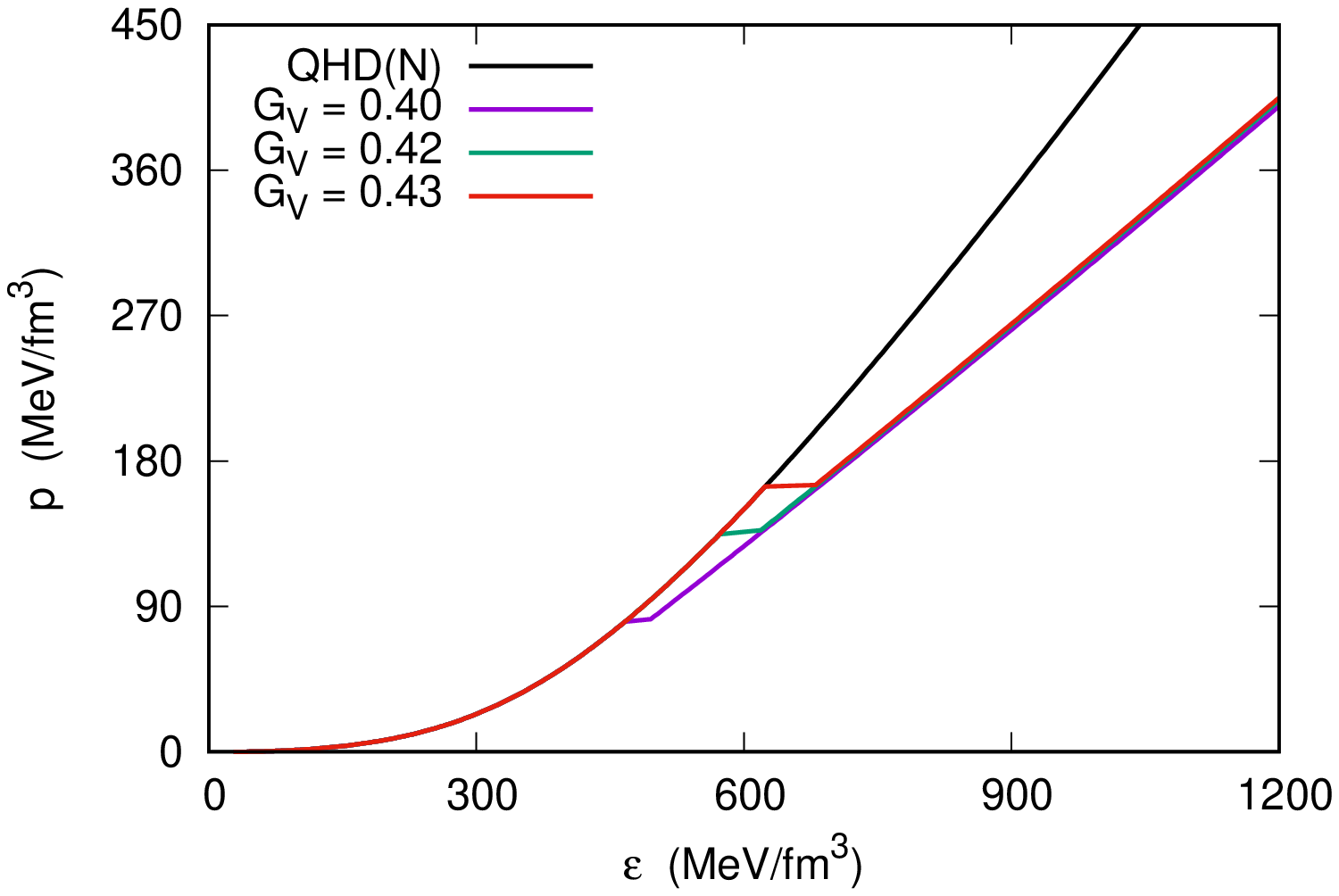} &
\includegraphics[width=0.33\textwidth,,angle=270]{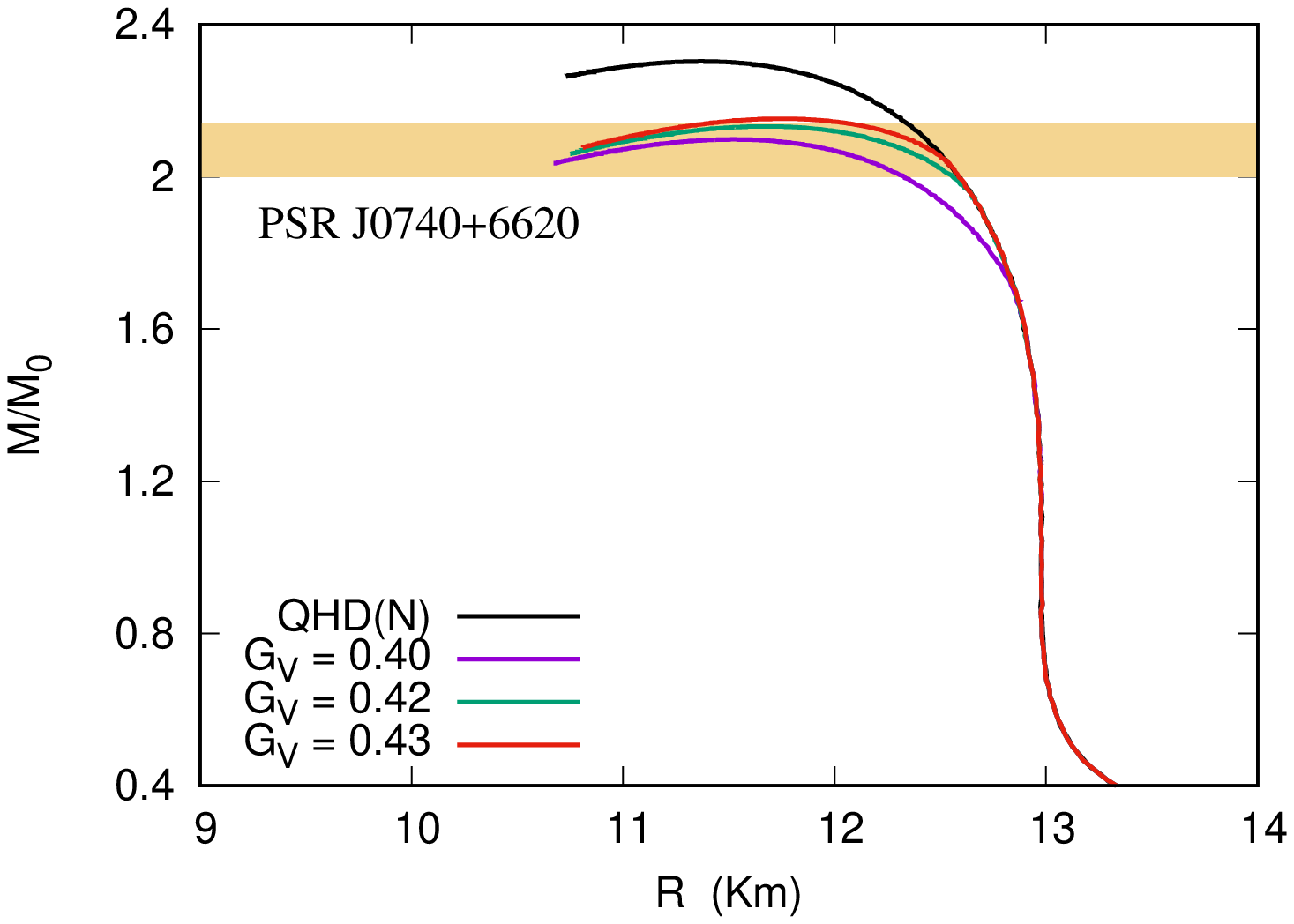} \\
\end{tabular}
\caption{(Color online) EoS (left) and TOV solution (right) for different values of $G_V$ without hyperons. A maximum mass of 2.15 M$_\odot$ is possible.} \label{F8}
\end{figure*}

\begin{center}
\begin{table}%[ht]
\begin{center}
\caption{Chemical potential and pressure at phase transition from hadron (H) to quark (Q) for different values of $G_V$ without hyperons. We also show the energy density (in MeV/fm$^3$) at both phases at the critical neutron chemical potential.}
\label{T9}
\scalebox{0.99}{
\begin{tabular}{|c|c|c|c|c|c|}
\hline 
  $G_V$ (fm$^2$)   &$\mu_n^H = \mu_n^Q$  & $p^H = p^Q$ & $\epsilon_H$ & $\epsilon_Q$ \\
 \hline
 0.40~ &  1206 MeV & 80 (MeV$/fm^3)$ & ~466~ & ~491~\\
 \hline
 0.42~ &  1319 MeV & 136 (MeV$/fm^3)$ & 572 & 614 \\
 \hline
 0.43~ &  1369 MeV  & 164 (MeV/$fm^3)$ & 415 & 440 \\
  \hline
  \end{tabular}}
\end{center}
\end{table}
\end{center}

\begin{table}%[ht]
\begin{center}
 \caption{Hybrid star properties for different values of the $G_V$ without hyperons.}
 \label{T10}
\begin{tabular}{|c|c|c|c|c|c|}
\hline
~$G_V$  (fm$^2$)& $M/M_\odot$ & $ R~(km)$ & $\epsilon_c$  & $M_{min}/M_\odot$  & $ R_{1.4}~(km)$  \\
\hline
  0.40   & 2.10  & 11.51  &  1196   &  1.67  & 12.96\\
 \hline
 0.42         & 2.13  & 11.70 &  1130   &  1.95  & 12.96\\
 \hline
0.43        & 2.15 & 11.75  & 1123   & 2.04 & 12.96  \\
\hline
QHD(N)        & 2.30 & 11.36  & ~1220~   & - & 12.96  \\
\hline
\end{tabular}
\end{center}
 \end{table}

In Fig.~\ref{F8} we display the EoS and the mass-radius relation obtained wiht the TOV solutions. The results are also presented in Table~\ref{T10}. We see that the energy gap here is even smaller than in the hyperonic case. Also, the gap grows with the increase of $G_V$. We also see that all models are able to reproduce the PSR J0740+6620 constraint. As in the case of hyperonic matter, the maximum mass is not very sensitive to the $G_V$ value. Here, the maximum mass grows 0.05 $M_\odot$,  only $2\%$ approximately. There are some coincidences here, when compared with the hyperonic case. For instance, the maximum mass with hyperons but no quarks is 2.15 $M_\odot$, which is the same value of the maximum mass found with quarks but no hyperons. Even more impressive are the properties of the maximum mass with $G_V = 0.40$ fm$^2$. With and without hyperons, this value is 2.10 $M_\odot$. The radius of the maximum mass is also equal to 11.51 km in both cases. Even the central energy density differs only by 1 MeV/fm$^3$. This indicates that the quark EoS, over the hadron EoS, is the more important term in hybrid stars, a result also obtained with the NJL model~\cite{lopes2021broken}. The main difference here is the minimum mass which produces the hybrid branch. With hyperons we have $M_{min}$ = 1.99 $M_{\odot}$, while without hyperons we have $M_{\odot}$ = 1.67 $M_{\odot}$. This is easily explained by the differences in the critical chemical potentials.

\begin{center}
\begin{table}%[ht]
\begin{center}
\caption{Masses and radii of the quark core and their proportional contribution for the maximally massive star within different hybrid stars models without hyperons.}
\label{T11}
\scalebox{0.99}{
\begin{tabular}{|c|c|c|c|c|c|c|}
\hline 
 $G_V$  (fm$^2)$  &  ~0.40~ &  ~0.42~  &  ~0.43~    \\
 \hline
 $M_Q/M_\odot$  & 1.10  & 0.72 & 0.55    \\
 \hline
 R$_Q$  (km) & 7.48  & 6.21 & 5.60   \\
 \hline
 $\%$ $M_Q/M_{max}$  & 52$\%$  & 33$\%$ & 25$\%$    \\
 \hline
 $\%$  R$_Q$/R$_{total}$ & 65$\%$  & 53$\%$ & 48$\%$       \\
 \hline
\end{tabular}}
\end{center}
\end{table}
\end{center}

Now we estimate the size and mass of the quark core in hybrid stars without hyperons. The results are presented in Table~\ref{T11}. The quark core is still pretty significant, reaching over $50\%$ for $G_V$ = 0.40 fm$^2$. This value is even larger than what is predicted by the pQCD presented in ref.~\cite{annala2020evidence}. Increasing $G_V$ makes the quark core smaller, reaching $25\%$ for $G_V = 0.43$ fm$^2$. Nevertheless, in all cases we have a quark core with masses above 0.50 $M_\odot$ and a radius above 5 km.

%%%%%%%%%%%%%%%%%%%%%%%%%%%%%%%%%%%%%%%%%%%%%%%%%%%%%%%%%%%%%%%%%%%%%%%%%%%%

\section{Comparison with the NJL model} 

For the sake of completeness, we construct also a hybrid star with another quark model, the so-called Nambu Jona-Lasinio (NLJ) model~\cite{nambu1961dynamical}. Indeed, here we are able to make a double comparison. We not only compare the differences of the NJL model with the extended versions of the MIT model discussed in the previous section, but we also compare the differences between the hybrid stars with the parametrization of ref.~\cite{lopes2021hyperonic}, and with the traditional GM1~\cite{glendenning2000compact}, as presented in ref.~\cite{lopes2021broken}.

The NJL and the MIT bag models and its extensions are effective models with complementary features. For instance, the MIT bag model is able to explain the  confinement due to the bag, which introduces a vacuum pressure in the Lagrangian, while the NJL model lacks confinement. On the opposite side, the NJL model presents chiral symmetry, which is a symmetry predicted by the QCD, while it is exactly the Bag term that breaks the chiral symmetry in the MIT bag model. Another difference is the interaction. In the vector MIT, as in the QHD, the strong force is simulated by the exchange of massive mesons, therefore these models are renormalizable. In the NJL model, there is no mediator. The interaction is a direct quark-quark point-like scheme. This makes the NJL model a non-renormalizable model, and a cutoff is needed to obtain physical results.
To make a direct comparison with ref.~\cite{lopes2021broken}, we use here a SU(3) version of the NJL, whose Lagrangian includes a scalar, pseudo-scalar and the ’t Hooft six-fermion interaction. Its Lagrangian reads:

\begin{eqnarray}
 \mathcal{L}_{NJL} = \bar{\psi}_q[\gamma^\mu(i\partial_\mu - m_q)\psi_q] + \nonumber \\
 G_S\sum_1^8 [(\bar{\psi}_q\lambda_a\psi_q)^2 + (\bar{\psi}_q\gamma_5\lambda_a\psi_q)^2]  + \nonumber \\
 - K\{ \det[\bar{\psi_q}(1+\gamma_5)\psi_q +\det[\bar{\psi_q}(1 - \gamma_5)\psi_q]\} \label{e13}
\end{eqnarray}
where $\psi_q$ are the quark Dirac fields, with three flavors, $m_q$ = diag$(m_u, m_d, m_s)$ are the current quark masses, $\lambda_a$ are the eight Gell-Mann flavor matrices and $G_S$ and $K$ are dimensionfull coupling constants. Here, as in ref.~\cite{lopes2021broken}, we use the so-called HK parametrization~\cite{hatsuda1994qcd}. Another term that can be summed to the Lagrangian is a vector channel:

\begin{equation}
 \mathcal{L}_{NJLv} =  G_V(\bar{\psi}_q \gamma^\mu\psi_q)^2 ; \label{e14}
\end{equation}
where $G_V$ is treated as a free parameter. The detailed calculation of the EoS for the SU(3) NJL model with and without the vector term can be found in ref.~\cite{lopes2021broken,hatsuda1994qcd,buballa2005njl} and the references therein.

\begin{figure}%[ht] 
\begin{centering}
 \includegraphics[angle=270,
width=0.5\textwidth]{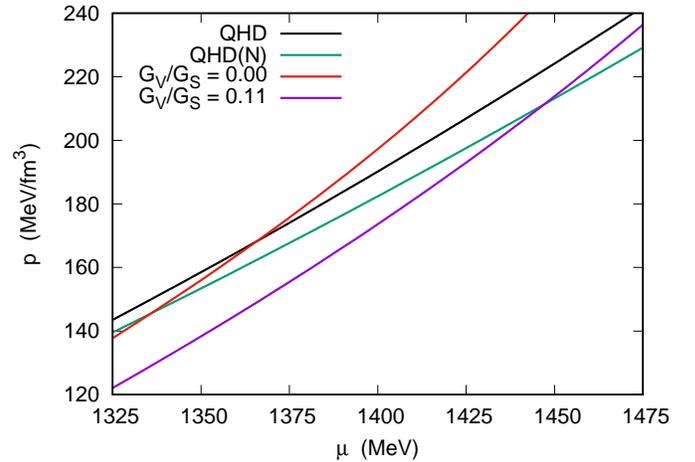}
\caption{(Color online) The pressure as a function of the chemical potential with and without $G_V$ for the NJL model and with and without hyperons for the QHD model.}
 \label{F9}
\end{centering}
\end{figure}

We plot in Fig.~\ref{F9} the pressure as a function of the chemical potential for the SU(3) NJL model with and without the vector term. Also we plot the QHD model with and without hyperons. The results are also displayed in Table~\ref{T12}. To differentiate models with and without hyperons in the QHD and with and without the vector term in the NJL model, we use the following scheme: A indicates QHD with hyperons and B indicates QHD without hyperons. The number 0 indicates $G_V/G_S = 0.00$ and the number 1 indicates $G_V/G_S$ = 0.11.

\begin{center}
\begin{table}%[ht]
\begin{center}

\caption{Chemical potential and pressure at phase transition from hadron (H) to quark (Q) for the above discussed models of the QHD and the NJL. We also show the energy density (in MeV/fm$^3$) at both phases at the critical neutron chemical potential. Here $A$ indicates QHD with hyperons and $B$ indicates QHD without hyperons. The number 0 indicates $G_V/G_S = 0.00$ and the number 1 indicates $G_V/G_S$ = 0.11.}
\label{T12}
\scalebox{0.99}{
\begin{tabular}{|c|c|c|c|c|c|}
\hline 
 -  &$\mu_n^H = \mu_n^Q$  & $p^H = p^Q$ & $\epsilon_H$ & $\epsilon_Q$ \\
 \hline
 B1~ &  1446 MeV & 210 (MeV$/fm^3)$ & ~699~ & ~1029~\\
 \hline
 B0~ &  1332 MeV & 143 (MeV$/fm^3)$ & 588 & 817 \\
 \hline
 A0~ &  1364 MeV  & 167 (MeV/$fm^3)$ & 682 & 921 \\
  \hline
  \end{tabular}}
\end{center}
\end{table}
\end{center}

As can be seen, for the hyperonic QHD model, only the NJL without the vector channel allows phase transtion. 
Nevertheless, even for the pure nucleonic EoS, for the NJL with the vector channel, the critical chemical potential is already above  1400 MeV. Also, we can notice, that for the HK parametrization, the critical chemical potential is always very high, above 1300 MeV. This fact contrasts with the MIT bag models, where, in its original form, it produces a critical chemical potential of only 968 for a Bag pressure value of $B^{1/4}$ = 165 MeV. When we compare the parametrization of the QHD used in this work with the GM1 used in ref.~\cite{lopes2021broken}, we see that the critical chemical potential for the GM1 is always smaller for the same value of $\alpha_v$ = 0.50: 1285 MeV (GM1)
vs 1364 MeV (here) without the vector term and 1437 MeV (GM1) vs 1660 MeV (here) for $G_V/G_S$ = 0.11.

\begin{figure*}%[ht]
\begin{tabular}{cc}
\includegraphics[width=0.33\textwidth,angle=270]{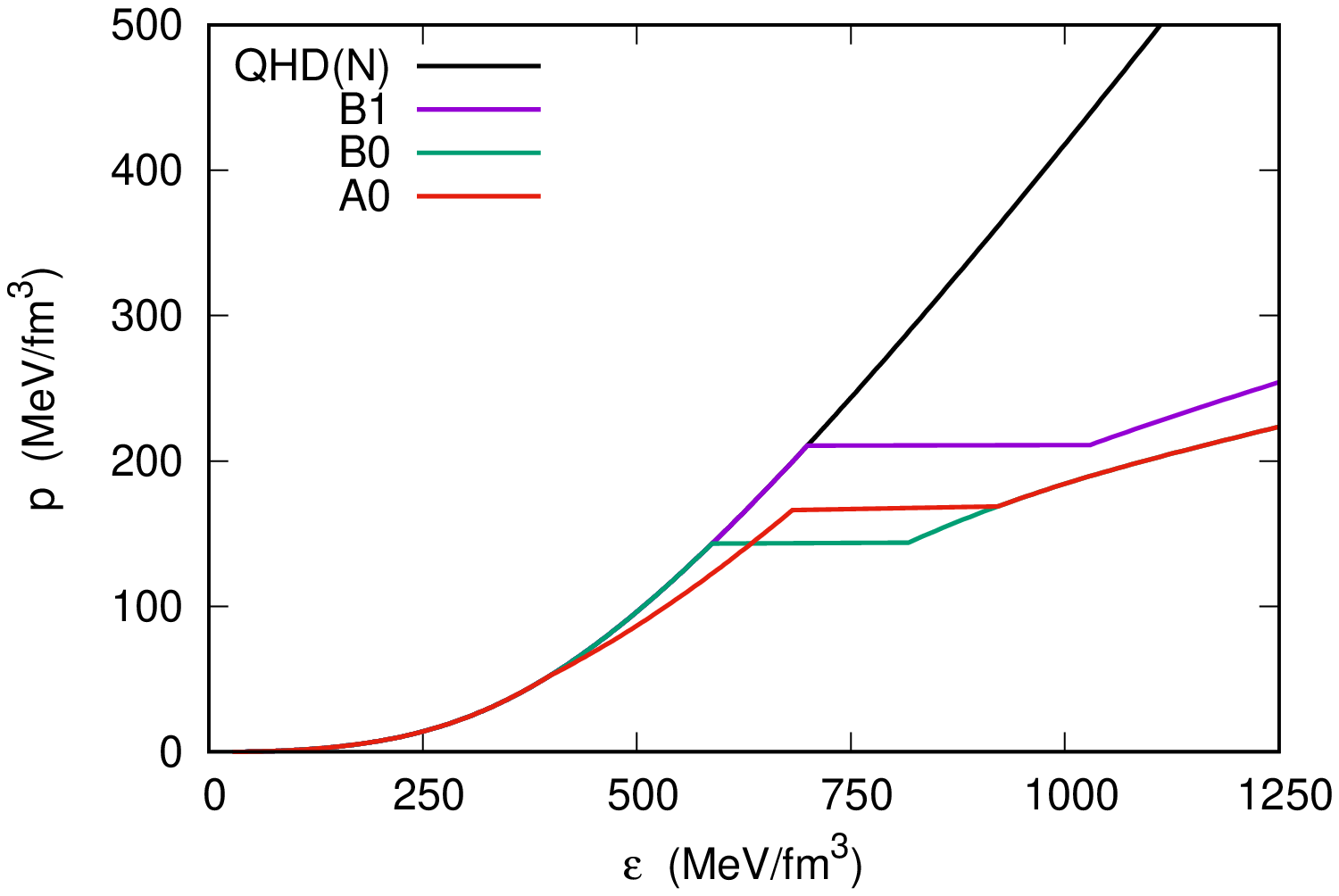} &
\includegraphics[width=0.33\textwidth,,angle=270]{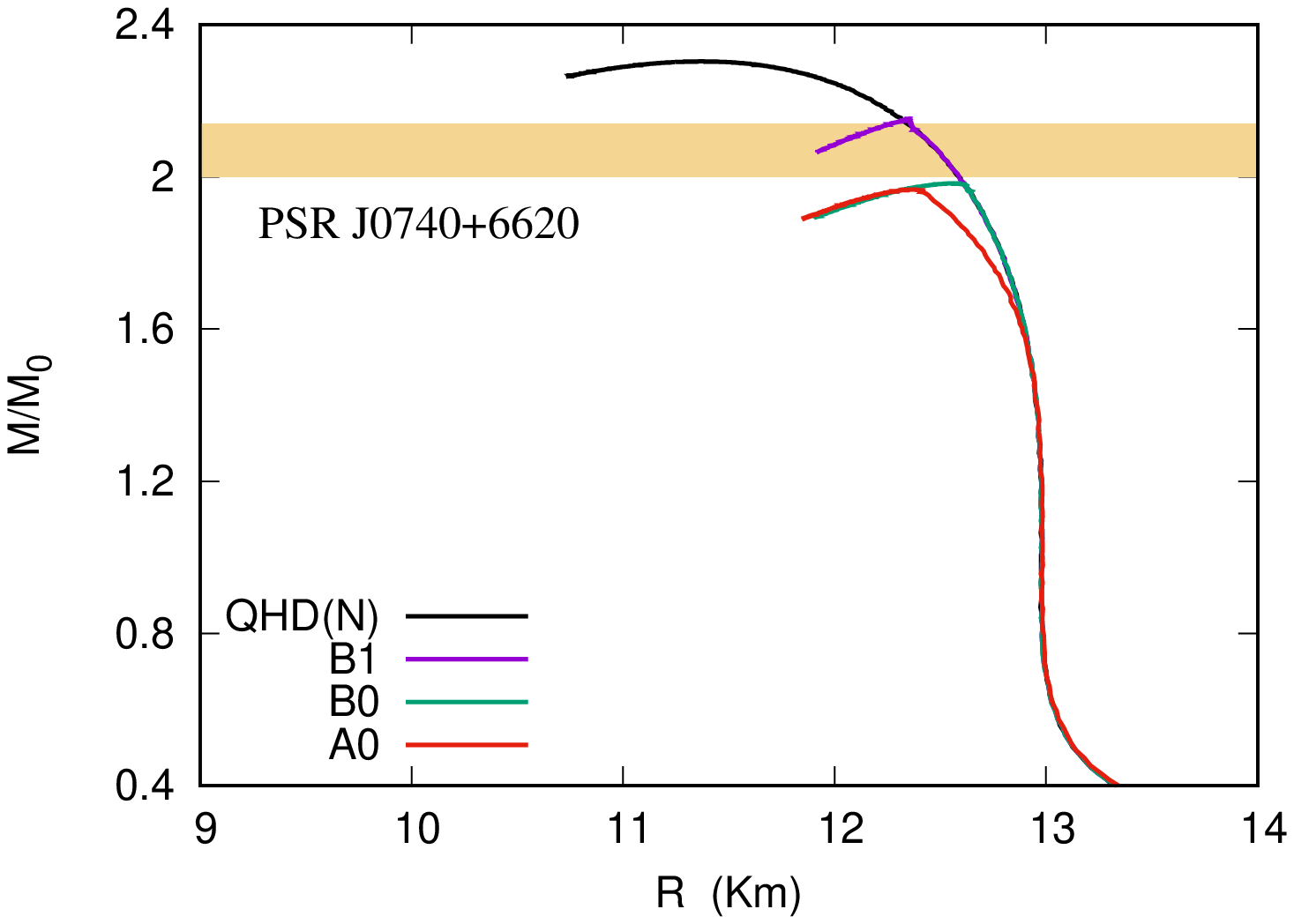} \\
\end{tabular}
\caption{(Color online) EoS (left) and TOV solution (right) for different values of $G_V$ for the NJL model and with and without hyperons for the QHD model. Here $A$ indicates QHD with hyperons and $B$ indicates QHD without hyperons. The number 0 indicates $G_V/G_S = 0.00$ and the number 1 indicates $G_V/G_S$ = 0.11.} \label{F10}
\end{figure*}

\begin{table}%[ht]
\begin{center}
\caption{Hybrid stars properties for different values of the $G_V/G_S$, with and without hyperons.}
\label{T13}
\begin{tabular}{|c|c|c|c|c|c|}
\hline
~ - & $M/M_\odot$ & $ R~(km)$ & $\epsilon_c$  & $M_{min}/M_\odot$  & $ R_{1.4}~(km)$  \\
\hline
  B1  & 2.15  & 12.35  &  947   &  12.15  & 12.96\\
 \hline
 B0         & 1.98  & 12.56 &  888   &  1.96  & 12.96\\
 \hline
A0        & 1.96 & 12.37  & 767   & 2.04 & 12.96  \\
\hline
QHD(N)        & 2.30 & 11.36  & ~1220~   & - & 12.96  \\
\hline
\end{tabular}
\end{center}
\end{table}

We display in Fig.~\ref{F10} the EoS and the mass-radius relation via TOV solutions for different configurations of the QHD and the NJL model. We see that the gap in the energy density is significantly larger in the NJL than in the vector MIT bag model in both configurations: with and without the Dirac sea contribution. We also see that the parametrization from Table~\ref{TL1} predicts a significantly larger gap when compared with the GM1 model~\cite{lopes2021broken}.

When we look at the mass-radius relation and in Table ~\ref{T13}, it seems that only the B1 configuration is able to fulfill the constraints of the PSR J070+6620, but it is not the case. When we compare  Table ~\ref{T12} with Table ~\ref{T13}, we see that only the B0 configuration produces a stable hybrid branch. In all other configurations, the maximum mass falls in the gap region. This is in contrast with the vector MIT bag model with and without the Dirac sea contribution, where always a stable hybrid branch is formed. This is also in contrast with the results within GM1 model, where most (although not all) parametrizations predict a stable branch~\cite{lopes2021broken}.
 
 \begin{center}
\begin{table}%[ht]
\begin{center}
\caption{Masses and radii of the quark core and their proportional contribution for the maximally massive star. Within NJL, only the B0 configuration produces stable hybrid branches.}
\label{T14}
\scalebox{0.99}{
\begin{tabular}{|c|c|}
\hline 
 -  &  ~B0~    \\
 \hline
 $M_Q/M_\odot$  & 0.031    \\
 \hline
 R$_Q$  (km) & 2.13   \\
 \hline
 $\%$ $M_Q/M_{max}$  & 1.55$\%$     \\
 \hline
 $\%$  R$_Q$/R$_{total}$ & 17$\%$       \\
 \hline
\end{tabular}}
\end{center}
\end{table}
\end{center}

We estimate next the quark core size and mass for the B0 configuration, despite the fact that it does not fulfill the PSR J0740+6620 constraint, and display the results in  Table~\ref{T14}. We can see that the results are very timid. The quark core mass does not reach $2\%$, in total contrast with the results obtained with the extended versions of the MIT bag model where the quark core mass varies from 23$\%$ to $81\%$. On the other hand, the results are just a little smaller then those presented in ref.~\cite{lopes2021broken} for the GM1 parametrization, where the quark core mass represents 3.3$\%$ 
(for $\alpha_v$ = 0.50).

As we have done in Sec.~\ref{s3}, we now compare the results for the NJL with other results presented in the literature. In ref.~\cite{ranea2016constant}, the authors use the same GM1 for the hadron phase, but with different hyperon-meson coupling constants; and use the same HK parametrization for the NJL model with and without the vector term. However, instead of using the Maxwell construction as the criteria for hadron-quark phase transition, they use the so called CSS parametrization, which produces a critical speed of sound, instead of a critical chemical potential. They found that for vector channel varying form 0.00 $<~G_V/G_S~<$ 0.30 the phase transition is always possible. 
However, the quark core masses are even smaller than what we found in the present work, their values varying from $10^{-5}~< M_Q/M_\odot~< 0.011$.

%%%%%%%%%%%%%%%%%%%%%%%%%%%%%%%%%%%%%%%%%%%%%%%%%%%%%%%%%%%%%%%%%%%%%%%%%%%%%

\section{The dimensionless tidal parameter and the speed of sound}

Before we finish our work, we analyze two important physical quantities: the first one is the dimensionless tidal parameter $\Lambda$, and compare it to the recent constraints presented in the literature coming from LIGO/VIRGO gravitational wave observatories and the GW170817 event~\cite{abbott2018gw170817}; and the second one is the square of the speed of sound, $v_s^2$, and its relation with the size and mass of the quark core in hybrid stars as presented in ref.~\cite{annala2020evidence}. As the models that produce higher maximum masses, as well as larger values of quark core masses and radii, are the models within the vector MIT bag model without the Dirac sea contribution, we restrict our analyzes here to the parameters presented in Table~\ref{T5}.

If we put an extended body in an inhomogeneous external field it will experience different forces throughout its extent. This is called a tidal interaction. The tidal deformability of a compact object is a single parameter $\lambda$, 
%(sometimes called Love parameter) 
that quantifies how easily the object is deformed when subject to an external tidal field. A larger tidal deformability indicates that the object is easily deformable. On the opposite side, a compact object with a smaller tidal deformability parameter is more compact and it is more difficult to deform it.

The tidal deformability is defined as the ratio between the induced quadrupole $Q_{i,j}$ and the tidal field $\mathcal{E}_{i,j}$ that causes the perturbation:

\begin{equation}
 \lambda =  - \frac{Q_{i,k}}{\mathcal{E}_{i,j}} . \label{e15}
\end{equation}

\begin{figure}%[ht] 
\begin{centering}
\includegraphics[angle=270,width=0.5\textwidth]{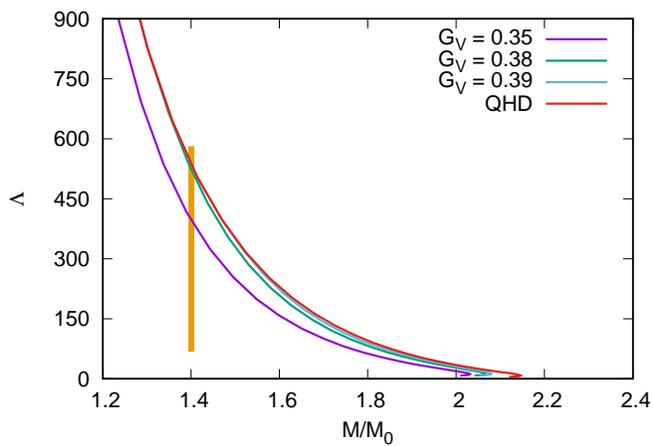}
\caption{(Color online) Tidal dimensionless parameter as a function of the star mass for different values of $G_V$ as presented in Table~\ref{T5}. $G_V$ = 0.40 $fm^2$ is not shown as its curve is almost identical to the QHD model. } \label{F11}
\end{centering}
\end{figure}

However, in the literature, the dimensionless tidal deformability parameter ($\Lambda$) 
is more commonly found and it is defined as:

\begin{equation}
 \Lambda~\dot{=}~\frac{\lambda}{M^5} = \frac{2k_2}{3C^5},
\end{equation}
where $M$ is the compact object mass and $C = GM/R$ is its compactness. The parameter $k_2$ is called Love number and is related to the metric perturbation. A complete discussion about the Love number and its calculation is both very extensive and well documented in the literature. Therefore, it is out of the scope of the present work. We refer the interested reader to~\cite{abbott2018gw170817,flores2020gravitational,chatziioannou2020neutron,hinderer2008tidal} and to the references therein. We plot the tidal dimensionless parameter for three different values of $G_V$, as well as for the pure hadronic model with hyperons (QHD) in Fig.~\ref{F11}, and show its value for the canonical mass ($\Lambda_{1.4}$) in Table~\ref{T15}. 

In ref.~\cite{abbott2018gw170817}, the authors constraint the tidal parameter for the canonical mass in the range between 70 $<~\Lambda_{1.4}~<$ 580. As can be seen, all our models fulfill this constraint. This is a direct consequence of the chosen QHD parametrization. Also, for $G_V$ = 0.35 fm$^2$, the tidal parameter is significantly smaller, as it also has a smaller radius for the canonical star. However, this is not the main reason, as pointed out in ref.~\cite{lopes2021modified-partI,hinderer2008tidal}, the tidal parameter is, in fact, more dependent of the microscopic $k_2$ than of the macroscopic compactness. For $G_V$ = 0.39 fm$^2$ and above, the $\Lambda_{1.4}$ is already equal to a pure hadronic EoS, as the minimum mass that corresponds to a hybrid star with this parametrization is already above 1.40 $M_\odot$.

\begin{center}
\begin{table}%[ht]
\begin{center}
\caption{Tidal dimensionless parameter for the canonical 1.4 $M_\odot$ star and the speed of sound of quark matter at the critical chemical potential.}
\label{T15}
\scalebox{0.99}{
\begin{tabular}{|c|c|c|c|c|c|c|c|c|}
\hline 
 $G_V$  (fm$^2)$  &  ~0.35~ &  ~0.38~  &  ~0.39~&~0.40~&    QHD  \\
 \hline
$\Lambda_{1.4}$  & 395  & 520 & 534 & 534 & 534   \\
 \hline
 $v_s^2$  & 0.40  & 0.42 & 0.42 & 0.45 & -   \\
 \hline
\end{tabular}}
\end{center}
\end{table}
\end{center} 

Now we turn our attention to the speed of sound. As pointed out in ref.~\cite{annala2020evidence}, the speed of sound of the quark matter is closely related to the mass and radius of the quark core in hybrid stars. The authors found that, in their own words: `if the conformal bound ( $v_s^2~  <$ 1/3 ) is not strongly violated, massive neutron stars are predicted to have sizable quark-matter cores'. Indeed, for speed of sound that satisfies the conformal bound, a quark core mass above 0.8 $M_\odot$ arises (see Fig. 6 from ref.~\cite{annala2020evidence}). We then calculate the square of the speed of sound of the quark phase and display the corresponding value at the neutron critical chemical potential for different values of $G_V$ in Table~\ref{T15}.

As can be seen, indeed, the higher the value of the speed of sound, the smaller the quark core in the hybrid star. However, unlike ref.~\cite{annala2020evidence}, the results here are not so drastic. For instance, for all the values of $G_V$ we have a speed of sound that violates the conformal limit. Moreover, while the differences in the speed of sound are about 10$\%$, the quark core mass differences reach over 100$\%$. Nevertheless, we agree with ref.~\cite{annala2020evidence} about the fact that a quark core in massive neutron star is not only possible, but probable.

%%%%%%%%%%%%%%%%%%%%%%%%%%%%%%%%%%%%%%%%%%%%%%%%%%%%%%%%%%%%%%%%%%%%%%%%%%%

\section{Conclusions}

In this work we studied hybrid stars, with the quark phase described by the extended versions of the MIT bag model via Maxwell construction. Using the QHD parametrization from ref.~\cite{lopes2021hyperonic} we were able to construct hybrid stars with masses above 2.0$M_\odot$ with a quark core reaching more than 1.6 $M_\odot$, while the dimensionless tidal parameter for the canonical mass still obeys $\Lambda_{1.4}~<$ 580. The main remarks are resumed below.

\begin{itemize}
 
 \item With the chosen QHD parametrization for hadronic matter and the original MIT bag model for quark matter, no stable hybrid star reaches the 
 PSR J07040+6620 inferior mass limit observation~\cite{riley2021nicer}.
 
 \item Within the vector MIT bag model, we were able to produce very massive hybrid stars, reaching 2.10 $M_\odot$, whose quark core mass and radius can both account for more than 80$\%$ of the total quantities of the hybrid star.
 
 \item The value of the vector channel $G_V$ has little influence on the maximum mass of the hybrid star, but has strong influence on the size of the quark core.
 
 \item When we took a term that mimics the Dirac quark sea contribution into account, the maximum mass became below the inferior limit of the PSR J07040+6620. Unlike the previous case, the $b_4$ parameter influences both, the size of the core and the maximum mass.
 
 \item Despite the fact that hyperons seem inevitable, as pointed out in ref.~\cite{lopes2021broken,gerstung2020hyperon}, we studied hybrid stars with no hyperon content and were able to produce even more massive hybrid stars, reaching 2.15 $M_\odot$.
 
 \item For $G_V$ = 0.40$fm^2$, the maximum mass and the corresponding radius of the hybrid star are independent of the presence or absence of hyperons. The main differences are the critical chemical potential and the minimum hybrid star mass, $M_{min}$. This indicates that is the quark EoS, over the hadronic one, that determines the maximum mass of a hybrid star. Similar results have already been discussed for the NJL model~\cite{lopes2021broken}.
 
 \item Compared with the NJL model, we saw that the vector MIT bag model produces more massive hybrid stars and larger quark cores. Moreover, within the chosen QHD parametrization combined with the NJL model, only the configuration without hyperons can produce a stable hybrid branch.
 
 \item For low values of the critical chemical potential, the radius of the canonical star and the corresponding tidal parameter are significantly reduced.
 
 \item We did not found a drastic dependence of the quark core on the speed of sound as suggested in ref.~\cite{annala2020evidence}. Indeed, we were able to produce hypermassive quark cores still violating the so-called conformal limit ($v_s^2~<$ 1/3). Nevertheless, we agree with the authors about the nature of massive compact objects being hybrid stars.
 
\end{itemize}

%%%%%%%%%%%%%%%%%%%%%%%%%%%%%%%%%%%%%%%%%%%%%%%%%%%%%%%%%%%%%%%%%%%%%%%%%%%%

\section*{Acknowledgements}

This work is a part of the project INCT-FNA Proc. No. 464898/2014-5. D.P.M. was partially supported by Conselho Nacional de Desenvolvimento Científico e Tecnológico (CNPq/Brazil) under grant 301155.2017-8  and  C.B. acknowledges a doctorate scholarship from Coordenação de Aperfeiçoamento de Pessoal do Ensino Superior (Capes/Brazil).

%%%%%%%%%%%%%%%%%%%%%%%%%%%%%%%%%%%%%%%%%%%%%%%%%%%%%%%%%%%%%%%%%%%%%%%%
\section*{Data Availability}

All data that support the findings of this study are included within the article (and any supplementary files).

% The best way to enter references is to use BibTeX:

%\bibliographystyle{mnras}
\bibliography{references} 

% Alternatively you could enter them by hand, like this:
% This method is tedious and prone to error if you have lots of references
%\begin{thebibliography}{99}
%\bibitem[\protect\citeauthoryear{Author}{2012}]{Author2012}
%Author A.~N., 2013, Journal of Improbable Astronomy, 1, 1
%\bibitem[\protect\citeauthoryear{Others}{2013}]{Others2013}
%Others S., 2012, Journal of Interesting Stuff, 17, 198
%\end{thebibliography}

% Don't change these lines
%\bsp	% typesetting comment
%\label{lastpage}
\end{document}